\documentclass[aps,prb,twocolumn]{revtex4}
\usepackage{etoolbox}
\makeatletter
\patchcmd{\NAT@citesuper}
   {\textsuperscript{#1}}{\textsuperscript{[#1]}}{}{}
\makeatother

\usepackage{natbib}
\setcitestyle{super}

\usepackage{epstopdf}

\usepackage{amsmath}
\usepackage{graphicx}
\usepackage{verbatim}
\usepackage{dcolumn}
\usepackage{amsmath}
\usepackage{latexsym}
\usepackage{MnSymbol}
\usepackage{textcomp}
\usepackage[thinlines]{easytable}
\newcommand{\RNum}[1]{\uppercase\expandafter{\romannumeral #1\relax}}
\makeatletter
\def\NAT@def@citea{\def\@citea{\NAT@separator}}
\makeatother
\usepackage{bm}
\newcommand*{\citen}[1]{%
  \begingroup
    \romannumeral-`\x 
    \setcitestyle{numbers,square}%
    \cite{#1}%
  \endgroup   
}

\begin{document}
\title{A high-performance MoS$_2$ synaptic device with floating gate engineering for Neuromorphic Computing}

\author{Tathagata Paul,$^1$\footnotemark[3] Tanweer Ahmed,$^1$ Krishna Kanhaiya Tiwari,$^2$ Chetan Singh Thakur$^3$ and Arindam Ghosh$^{1,4}$\footnote[3]{e-mail:tathagata@iisc.ac.in, arindam@iisc.ac.in}}

\address{$^1$Department of Physics, Indian Institute of Science, Bangalore 560012,India.  $^2$Visva Bharati University Santiniketan, West Bengal 731235, India. $^3$Department of Electronic Systems Engineering, Indian Institute of Science, Bangalore 560012, India. $^4$Centre for Nanoscience and Engineering Indian Institute of Science,Bangalore 560012, India.}

\begin{abstract}
As one of the most important members of the two dimensional chalcogenide family, molybdenum disulphide (MoS$_2$) has played a fundamental role in the advancement of low dimensional electronic, optoelectronic and piezoelectric designs. Here, we demonstrate a new approach to solid state synaptic transistors using two dimensional MoS$_2$ floating gate memories. By using an extended floating gate architecture which allows the device to be operated at near-ideal subthreshold swing of 77 mV/decade over four decades of drain current, we have realised a charge tunneling based synaptic memory with performance comparable to the state of the art in neuromorphic designs. The device successfully demonstrates various features of a biological synapse, including pulsed potentiation and relaxation of channel conductance, as well as spike time dependent plasticity (STDP). Our device returns excellent energy efficiency figures and provides a robust platform based on ultrathin two dimensional nanosheets for future neuromorphic applications.

\end{abstract}

\maketitle

 Understanding the complexities in the functioning of the human brain has been one of the foremost challenges in the field of neuroscience. Among the several proposed models, only a few can explain the operation of a human brain and that too for a very limited set of functionalities~\cite{hebb_book,guillery_binocular,miller_synaptic}. From an electronic point of view, the computational architecture of a brain is vastly different from that of a traditional von Neumann architecture based system~\cite{Neumann_computation,hennessy_book}. This has led to the emergence of neuromorphic computation schemes~\cite{Mead_neuromorphic,thakur_large,thakur_analogue,thakur_neuromorphic,thakur_sound}. Current computation follows an architecture where processing and storage of data is handled by separate entities whereas in neuromorphic computation, processing and storage of data is handled by a single element which acts as the electrical analogue of a synapse. Mimicing the functionality and density of synapses in the brain would lead to a massive reduction in energy consumption  and immensely enhance computational capabilities like parallel processing. Given the high density of synapses required, traditional silicon based devices which are plagued by power dissipation and short channel effects are rendered unsuitable for scalable neuromorphic applications~\cite{young_short,desai_mos2}. This makes ultrathin two dimensional materials a perfect candidate for the active element of a synaptic transistor given their immunity to short channel effects and excellent gate coupling at nanometer length scales~\cite{desai_mos2,Peide_Ye_Channel_length_scaling}. 

Biologically, a synapse functions by changing its conductivity based on the sequence of synaptic pulses it receives. This is accomplished by varying the concentration of neurotransmitters or chemical stimulants which control the conductivity of the junction between two neurons~\cite{lodish_neurotransmitters}. An ideal synaptic transistor must  possess the twin qualities of being a non-volatile memory while inculcating a learning based mechanism to deduce its conductance from the history of applied inputs~\cite{zhu_ion,shi_correlated,yang_all,sangwan_multi,tian_anisotropic,yang_synaptic,chang_short,jo_nanoscale,lai_ionic,zhu_artificial,gkoupidenis_neuromorphic,kim_carbon,balakrishna_nanoionics,xu_organic,van_non,yan_graphene}. A considerable amount of literature currently exists on transition metal oxide based synaptic devices in both two terminal memristor and three terminal transistor geometry~\cite{yang_all,yang_synaptic}. However, oxides in general have a large band gap and require ionic liquid gating which diminishes the long term usability of these devices because of the short lifetime of most liquid gates. Furthermore, most of these devices utilise some form of electrochemical reaction to alter the concentration of an ionic species, and hence the channel conductance, making them very sensitive to environmental conditions like humidity, temperature etc.~\cite{yang_synaptic}. The requirement of a liquid gate can be avoided by substituting the transition metal oxide with a chalcogenide like molybdenum disulphide (MoS$_2$)  because of its comparatively lower band gap and better coupling to metallic gates \cite{desai_mos2}. MoS$_2$ has already been used as an active element in high quality non-volatile  memory cells with high ON/OFF ratio~\cite{Philip_Kim_controlled_charge_trapping,Andras_kis_MoS2_memory,Seongil_MoS2_memory, Choi_Mos2_memory} and appears to be a prime candidate for a complete solid state based synaptic transistor. It is a scalable semiconducting platform, with a layer dependent bandgap in the visible range~\cite{Andras_Kis_Nature_Nano_review,Jar_TMDC_review,Kallol_Gr_MoS2_hybrid_Nat_Nano}, exhibits a respectable carrier mobility (1-30~cm$^2$/Vs) and displays unique transport properties like variable range hopping, percolative switching and valleytronic effects~\cite{Subhamoy_ACS_Nano,Percolative_switching,Andras_Kis_MoS2_transistor,Dong_Sun_Valley,Ji_Feng_Valley,Tony_F_Heinz_Valley,cui_valley,Coupled_spin_valley_Wang}.

However, the current architecture of floating gate (FG)  memory with MoS$_2$ is not conducive for realistic neuromorphic applications as it needs large gate voltage pulses ($\sim$ 30~V) in three terminal geometry~\cite{Philip_Kim_controlled_charge_trapping} while a large energy dissipation per pulse is observed when the device is operated in two terminal mode~\cite{vu_two}. In this paper, we have addressed this difficulty by adopting an extended FG device architecture for the MoS$_2$ FET. Owing to its two-dimensional nature, MoS$_2$ can be readily inserted in a planar floating gate (FG) architecture, where one or more metallic layers (the FGs) act as temporary storage of charge induced by a global back or top gate~\cite{Philip_Kim_controlled_charge_trapping,wang_floating,vu_two}. FG memory devices have been deployed in MOS architecture for a considerable period of time, where the tunneling of charge between the channel and the FG enables storage of information~\cite{frohman_electrically,diorio_single}. With improvements in fabrication techniques for two dimensional systems, it is possible to create a two dimensional analogue of a FG memory by stacking different van der Waal layered materials on top of each other in an atomic lego or heterostructure~\cite{geim_van}. We incorporate this idea in our work and demonstrate the performance of a floating gate memory device with MoS$_2$ as the active element. We have implemented an extended graphene FG in our devices enabling us to improve the gating efficiency which consequently leads to an almost ideal subthreshold swing and reduces the required drain bias and switching pulse for stable memory action. These benefits extend to neuromorphic applications leading to a reduction in the pulse heights required for long term potentiation and depression of the channel which reduces the stress on the gate dielectric while improving the integrability of the device with current neuromorphic systems. The FG and the channel are separated by a hexagonal boron nitride (hBN) tunnel barrier which controls the charge transfer between them, enabling us to tune the channel conductance. Distinct from previous reports of MoS$_2$ based synaptic memtransistors, which utilized bias induced motion of defect states in CVD (chemical vapour deposition) grown thin films to demonstrate the effect~\cite{sangwan_multi}, here we explore the possibility of controlled charge tunneling mediated multiple conductance states and synaptic activity in defect-free exfoliated MoS$_2$ layers. Using an extended FG architecture, we demonstrate hysteretic switching at near ideal subthreshold swing (77 mV/dec) in a trilayer stack of MoS$_2$, hBN and graphene. We establish quantitatively that the hysteresis is  caused by charge tunneling through hBN, and exploit the same to emulate spike time dependent plasticity at energy dissipation below~0.3~pJ.

\begin{figure*}[t]
\includegraphics[width=0.82\linewidth]{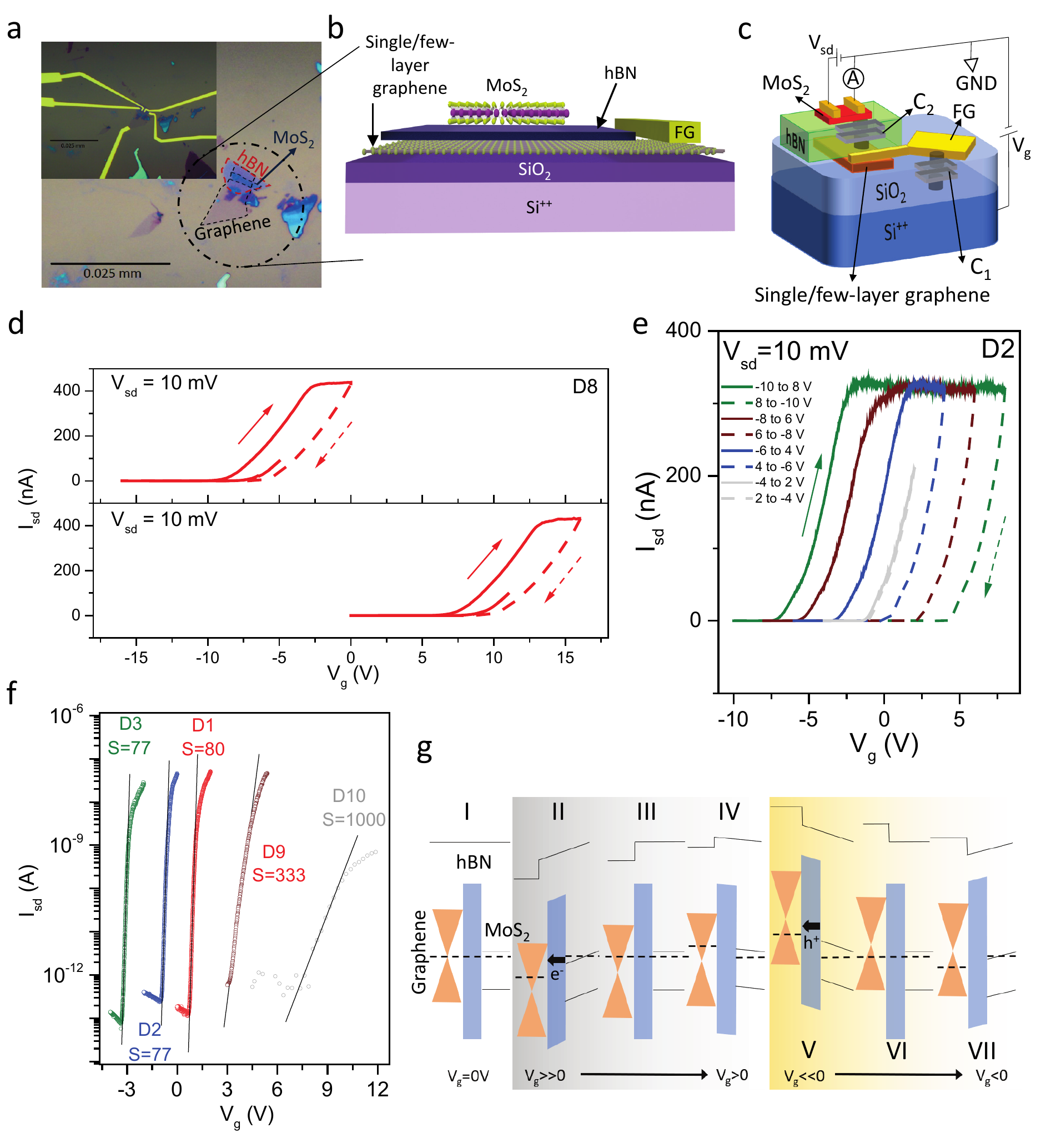}
\caption{Device structure and electrical characterisation. Optical micrograph of a typical device (a) and a schematic representation of the same (b). (c)  Representative image of the gate capacitance circuit. $C_1$~and~$C_2$ are the FG - Si$^{++}$ (across the SiO$_2$ dielectric) and FG - channel (hBN) capacitance respectively. FG is the large area metallic floating gate connected to the graphene layer. (d) Anti-hysteretic transfer characteristics  in extended floating gate devices. We can control the position of the hysteresis window by changing the center of the gate voltage sweep range. (e) Back gate sweep range dependence of observed antihysteresis. (f) Comparison of subthreshold slope for devices with different FG configurations. D1, D2 and D3 are devices with an extended FG, D9 has no extension of the FG and D10 is a device without a FG (device details in Supplementary Table~S1). The values of the subthreshold swing are mentioned in units of mV per decade beside the respective plots. The plots have been shifted horizontally for clarity. Transfer characteristics for D1, D2, D3 and D9 were performed at a $V_{sd}$ = 50~mV while that for D10 is obtained at $V_{sd}$ = 10~mV. (g) Schematic demonstrating the transport mechanism in the MoS$_2$ FG devices. Black arrows depict the direction of flow of charge during the potentiation and depression cycles between the FG and channel MoS$_2$. e$^-$ and h$^+$ denote electron and hole respectively. }
\end{figure*}

\begin{figure*}
\includegraphics[width=0.82\linewidth]{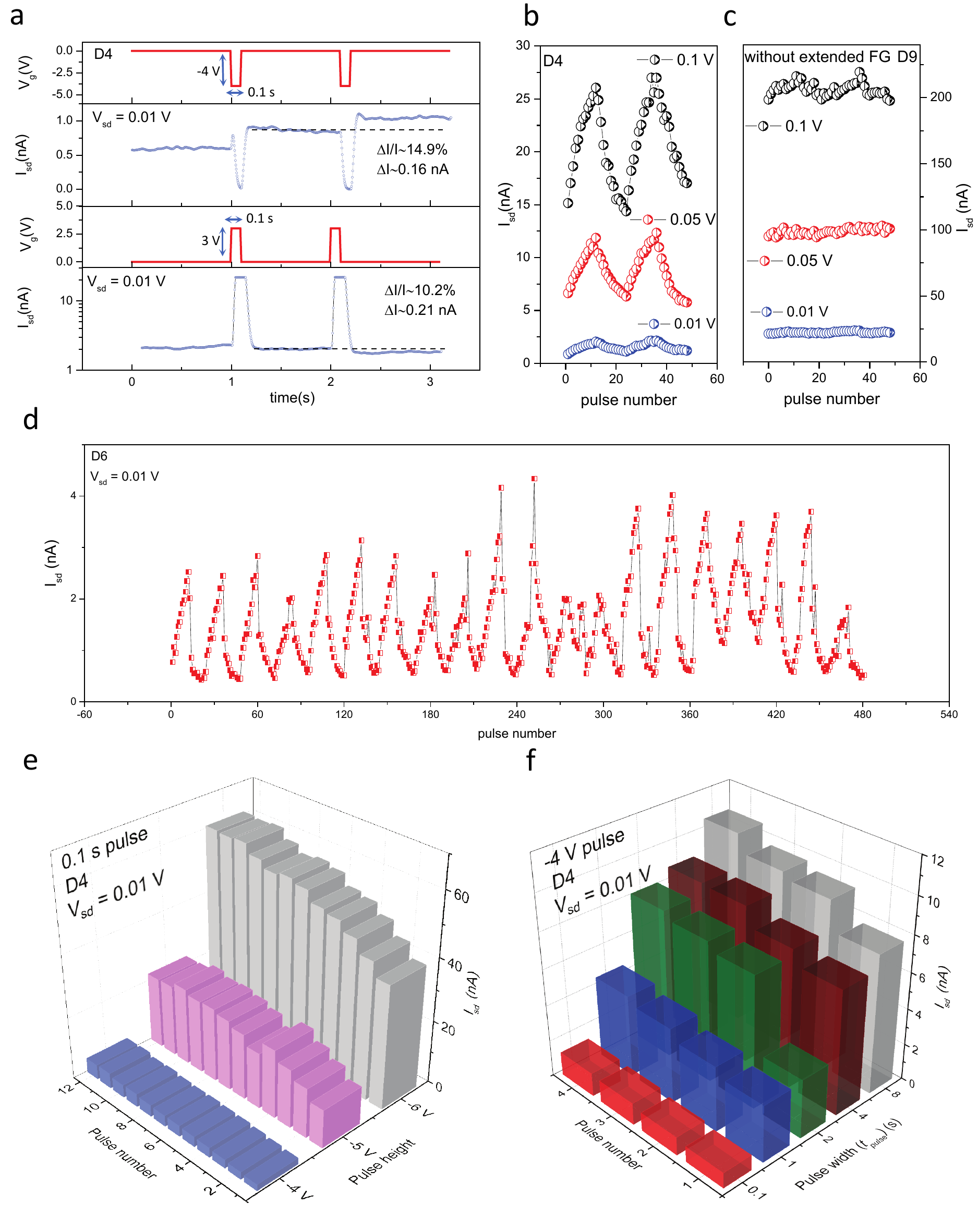}
\caption{Pulsed potentiation and depression in MoS$_2$ FG devices. (a) Time series data of drain current ($I_{sd}$) for potentiation (negative) and depression (positive) pulses. The absolute and percentage change in drain current is indicated in the respective sections. We use a pulse height of -4V  and +3V  for potentiation and depression respectively. Pulse width in both cases is 100 ms. The initial current values for potentiation (second panel from top) and depression (bottom panel) are different since the depression measurements were performed after a set of potentiation pulses had been applied which led to an increase in the channel conductance. Change in channel conductance for multiple potentiation and depression pulses for a device with (b) and without (c) an extended FG respectively. In the figures, pulses 1 to 12 and 25 to 36 are potentiation pulses while pulses 13 to 24 and 37 to 48 are depression pulses. Different potentiation and depression curves are obtained by varying the drain bias which is mentioned in volts beside the respective plots. Pulses used are similar to those in subsection (a) of this figure. (d) Repeatability of synaptic plasticity demonstrated for 20 cycles of potentiation (-3~V) and depression pulses (+3~V). Comparison of potentiation effect for different pulse heights at constant pulse width (e) and for different pulse widths at constant pulse height (f).   }
\end{figure*}


The experiments were performed on a heterostructure of mechanically exfoliated flakes of MoS$_2$, hBN and single/few layer graphene placed on a conventional $p^{++}$-Si/(285~nm)SiO$_2$ substrate~(Figure~1(a)) (details of devices used provided in Supplementary Table~S1). Individual layers were first exfoliated separately, searched under an optical microscope for suitable flakes using optical contrast and characterized by Raman spectroscopy for MoS$_2$ and graphene~(see Supplementary Figure~S1). The thickness of the hBN flake ($\approx$~5~nm - 7~nm) was obtained via AFM measurements (see Supplementary Figure~S2). We fabricated the heterostructure (Figure~1(b)) using a dry transfer method in an optical microscope with precision rotation and translation stages which assisted in the alignment of the individual layers~\cite{aamir20172d}. Electrical contacts were defined using electron beam lithography followed by metallization via thermal evaporation of Cr(5~nm)/Au(50~nm). The extended FG was fabricated by lithographically connecting the graphene layer to a large area floating gold pad as shown in Figure~1(c). The use of hBN as the intermediate layer was prompted by its excellent dielectric properties in the single crystalline form and large band gap ($\sim$~6~eV), which allows  a controlled charge tunneling while reducing  unintentional leakage of charge and providing a defect free substrate for the MoS$_2$ channel~\cite{Watanabe_hbn_bandgap,Subhamoy_APL_Mat,Dean_graphene_mobility,pari_noise_review,Pari_current_crowding}. Extension of the floating gate increases the total area of the SiO$_2$ capacitor ($\approx$~45000~$\mu$m$^2$, the area of the FG) which results in $C_1 \gg C_2$ in Figure~1(c), where $C_1$ and $C_2$ are the SiO$_2$ ($\approx$~5.72~pF) and hBN ($\approx$~5.9~fF) capacitance respectively. This increases the effective Si$^{++}$ - channel capacitance to that across the hBN layer only.

Figure~1(d) and (e) demonstrates the back gate transfer characteristics observed in the extended floating gate MoS$_2$ synaptic transistors. The threshold voltage is lower for the forward sweep (solid line) and higher for the reverse sweep (dashed line) leading to an anti-hysteretic transport. Additionally, the hysteresis window is sweep range dependent. We see a continuous decrease in the hysteresis window size (defined by the difference between the threshold voltage for the reverse and forward sweep) from $\approx$~11~V for the largest sweep range of $\approx$~18~V (Figure~1(e)) to a hysteresis-free transport for sweep ranges below $\approx$~6~V (Figure~1(e)). From Figure~1(d), we observe that the entire hysteresis window can be located at any chosen range of gate bias by changing the center of the back gate sweep range. The top and bottom panel in Figure~1(d) show hysteresis windows centered about a negative and positive gate bias, respectively, with one centered about zero gate bias depicted in Figure~1(e).  Hence, we see hysteresis even when the back gate voltage is either positive, negative or changing between positive and negative values during the sweep (Figure~1(d),(e)). This is important since a control over the threshold voltage is an essential component in designing a power efficient FET~\cite{Sakurai_alpha_power_law,Howritz_low_power_CMOS,Martin_transrerional_model}.
Figure~1(f) compares the subthreshold slope for five devices with varying configurations of the FG. Devices with an extended FG (D1, D2 and D3) demonstrate an almost ideal subthreshold slope of $\approx$ 80 mV/decade which increases to $\approx$ 300 mV/decade on removing the extension of the FG (D9) while devices with no FG (D10) operate at an even larger subthreshold slope of $\approx$ 1000 mV/decade. Capacitance engineering via extension of the FG leads to faster ON/OFF transitions with improved energy efficiency, both of which are of considerable importance in neuromorphic applications (Supplementary section \RNum{2}).

To explain the hysteresis in these devices, we postulate a charge trapping mechanism as shown in Figure~1(g). Starting from an initial flatband condition at zero back gate bias, $i.e.$ $V_g$=~0~V, we increase $V_g$ leading to an electron doping in MoS$_2$. Some electrons tunnel through the hBN into graphene (indicated by black arrow pointing in the  direction of charge transfer in Figure~1(g) (panel \RNum{2})) leading to a screening of the gate voltage as indicated in panel \RNum{2}. On decreasing the gate bias, this screening enables us to attain the flatband condition or OFF state at a value of $V_g>$~0~V (panel \RNum{3} of Figure~1(g)). Further decreasing the gate bias leads to a tunneling of electrons from graphene to the MoS$_2$ layer (or equivalently holes from the MoS$_2$ to graphene layer)~(panel \RNum{4} and \RNum{5} of Figure~1(g)). The positive charge on the graphene layer now screens the negative gate bias as shown in panel \RNum{5} of Figure~1(g). Like the positive bias condition, this screening leads to the flatband condition at an effective negative bias when we start the forward run resulting in the anti-hysteretic transfer characteristics. The sweep-rate independence (Supplementary Figure~S4) and range-tunability of the anti-hysteresis (Figure~1(d) and (e)) suggests (nearly) relaxation-free charge transfer between the channel and the FG which is facilitated by crystallinity of the hBN layer and atomically pristine van der Waals interfaces.

The plasticity of vertical charge transfer in the MoS$_2$ floating gate device allows non-volatile conductance change under pulsed gate operation.  This behaviour is analogous to biological synapses where the application of an excitatory or inhibitory pre-synaptic pulse has the effect of increasing or reducing the conductance of the synapse respectively. In this case, the gate acts as the pre-synaptic terminal and controls the conductance of the MoS$_2$ channel/synapse  using a sequence of pulses. The increase and decrease in conductance are known as potentiation and depression of the synapse respectively. This is performed by applying short time period (0.1~s) voltage pulses at the gate terminal while simultaneously tracking the change in drain current.  The channel conductance continuously increases for every excitatory pulse ($-4$~V pulse in top panel of Figure~2(a)) following an approximately linear pattern and decreases on application of an inhibitory pulse ($+3$~V pulse in the third panel from top of Figure~2(a)).  Figure~2(b) and (c)  compares the nature of synaptic response in devices with and without an extension of the FG respectively. Starting from the rest condition a set of twelve excitatory pulses ($-4$~V pulse height and 0.1~s pulse width) followed by twelve inhibitory ones ($+3$~V pulse height and 0.1~s pulse width) were applied at the gate terminal twice and the change in drain current was recorded after each pulse. The device with an extended FG (D4) shows a considerable change ($\lesssim$~80$\%$) in channel conductance (Figure~2(b)), while a negligible change is observed ($\lesssim$~2$\%$) in the device without an extended FG (D9) (Figure~2(c)). The current values plotted in Figure~2(b) and (c) shows the average current over a period of one second after the pre-synaptic pulse has been removed and the channel conductance has settled down to its final value (see Supplementary Section \RNum{9}). The long term plasticity is robust and persists even after a large number of potentiation and depression cycles which was limited to 20 in the current experiment (Figure~2(d)). We observe potentiation and depression curves similar to previously reported  synaptic devices~\cite{zhu_ion,shi_correlated,yang_all,sangwan_multi,tian_anisotropic,yang_synaptic,jo_nanoscale} although the shape of the excitatory post-synaptic current ($I_{sd}$ vs time plots in Figure~2(a)) in our case is different from that observed in previous reports ~\cite{zhu_ion,shi_correlated,yang_all,yang_synaptic,chang_short,jo_nanoscale}. As a result of the unique transport mechanism of these devices, we observe low conductance values during the time period of an excitatory (potentiation) pulse while higher values of conductance are seen during an inhibitory (depression) pulse (Figure~2(a)). Additionally, the inhibitory nature of positive gate voltage pulses leads to negative values for the short term plasticity based paired-pulse facilitation (PPF) index (see Supplementary Section \RNum{10} for details). We find that pulses of similar time period but larger magnitude produce a larger change in conductance. This is illustrated in Figure~2(e) for multiple potentiation cycles. A similar effect is observable on increasing the time period of the pulse while keeping the magnitude same (Figure~2(f)).

\begin{figure*}
\includegraphics[width=1\linewidth]{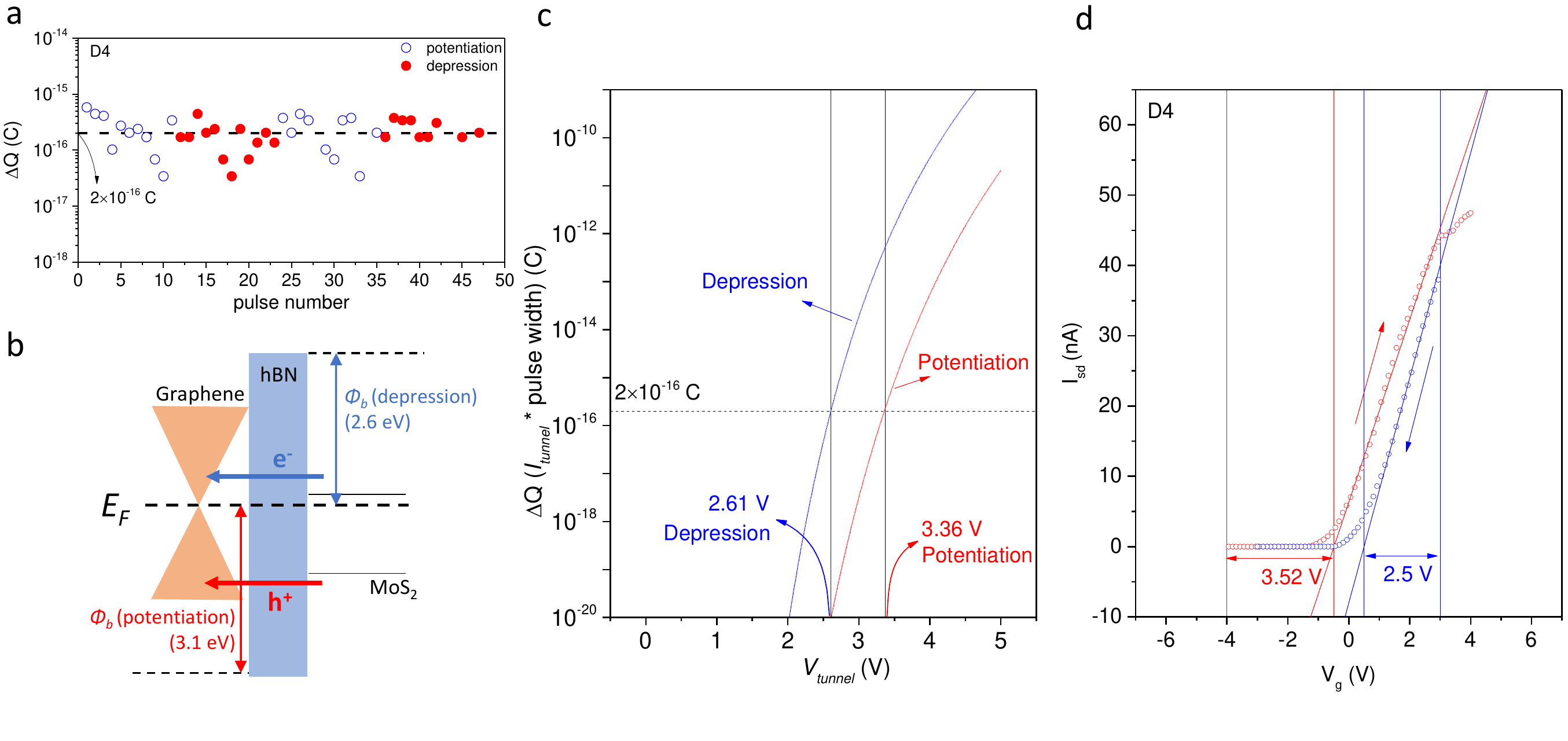}
\caption{Quantitative analysis of charge transport. (a) Graph depicting the tunneling charge as a function of pulse number for both potentiation and depression of channel conductance. (b) Schematic showing the tunnel barriers for tunneling of electrons ($\phi_b$ (depression)) and holes ($\phi_b$ (potentiation)). (c) Plot of the tunneling charge obtained using Fowler Nordheim theorem as a function of $V_{tunnel}$. The effective tunnel bias in the current device is obtained by finding the value of $V_{tunnel}$ corresponding to the average charge transferred per pulse. (d) Device transfer characteristics with markers indicating the difference between the threshold voltage and the extreme gate bias applied for both forward and reverse sweep directions.  }
\end{figure*}

For a quantitative analysis of the change in $I_{sd}$ during both potentiation and depression pulses, we consider the bi-directional tunneling of charge across the hBN layer. As discussed in Figure~1(g), the channel conductance varies due to the tunneling of charges in or out of the channel through a hBN tunnel barrier. In Figure~3(a), we plot the absolute value of charge transferred per excitatory ($-4$~V) or inhibitory ($+3$~V) pulse for the device D4. This is computed by finding the effective gate bias necessary to induce the change in drain current ($\Delta I_{sd}$) observed for a single potentiation/depression of the channel. The magnitude of charge exchanged during a potentiation or depression event can be estimated from $\Delta Q = \Delta V_g \times C_{self}$ where $\Delta V_g = 
\Delta I_{sd}/g_m$ with $g_m$ the transconductance and $\Delta V_g$ the effective change in gate voltage for a single pre-synaptic pulse. Here, $C_{self}$~($\approx$~8$\epsilon_{0}\sqrt{A_{FG}}$), $\epsilon_{0}$ and $A_{FG}$ are the self-capacitance of the FG, permittivity of free space and area ($\approx$~45000~$\mu$m$^2$) of the FG, respectively. The computed values of $\Delta Q$ for the potentiation and depression cycles depicted in Figure~2(b) ($V_{sd}$ = 0.01~V) are shown in Figure~3(a). We find $\Delta Q$ to be reasonably constant, being $\approx$~2$\times$10$^{-16}$~coulomb per pulse. To estimate the tunneling current ($I_{tunnel}$), we assume Fowler Nordheim type electric field dependent tunneling in our devices as reported previously~\cite{lee_electron} for hBN tunnel barriers. The tunneling current is given by
\begin{equation}
\label{Fowler_Nordheim}
   I_{tunnel}(V) =  \frac{A_{ch}q^3mV_{tunnel}^2}{8\pi h\phi_{b}d^2m^*}\exp[\frac{-8\pi \sqrt{2m^*}\phi_b^{\frac{3}{2}}d}{3hqV_{tunnel}}]
\end{equation}
where $A_{ch}$ is the channel area and $\phi_b$ the barrier height for tunneling.  The effective electron mass for hBN, $m^*=$~0.26$\times m$, where $m$ is the free electron mass.  Here, $h$ and $q$  represent the Plank's constant and electron charge, respectively, while $d\approx$ 5.8~nm is the thickness of the hBN layer (see Supplementary Figure~S2). The barrier height ($\phi_b$) is computed from the device band structure using known values for the work function of graphene and MoS$_2$ along with the electron affinity and band gap of hBN as shown in Figure~3(b)~\cite{Philip_Kim_controlled_charge_trapping}. We find a barrier height of 3.1~eV for potentiation which involves transfer of holes from MoS$_2$ to FG and 2.6~eV for depression which involves transfer of electrons (Figure~3(b)). In Figure~3(c) we have plotted the tunneling charge ($I_{tunnel} \times$ pulse width), calculated from Eq.~\ref{Fowler_Nordheim} for both potentiation and depression as a function of the tunneling bias ($V_{tunnel}$). $V_{tunnel}$ for the current devices are obtained by graphically solving  Eq.~\ref{Fowler_Nordheim} for known values of the tunneling charge from Figure~3(a), which yields the potential across the hBN layer to be 3.36~V and 2.61~V for potentiation and depression events respectively (Figure~3(c)). To verify this, we measure the effective bias across the hBN tunnel barrier (denoted by the difference in Fermi level between the graphene and MoS$_2$ layers in panel \RNum{5} (potentiation)  \RNum{2} (depression) in Figure~1(g)) from the device transfer characteristics (Figure~3(d)). The tunneling voltage for potentiation (depression) is given by the difference between the threshold voltage for forward (reverse) sweep and the excitatory (inhibitory) pulse height. This method yields $V_{tunnel}$ values of 3.52~V for potentiation and 2.5~V for depression (Figure~3(d)), which are similar to those obtained from Fowler Nordheim modelling (Figure~3(c)), confirming the charge tunneling mediated synaptic behaviour in our devices.  Since the synaptic activity originates from the tunneling of charges between the channel and the FG, we also observe synaptic plasticity in a two terminal geometry. However, the device operates at large current levels ($\approx$ few $\mu$A) making it  energetically unfavorable for neuromorphic applications (see Supplementary Section \RNum{8} for details).

\begin{figure}
\includegraphics[width=1\linewidth]{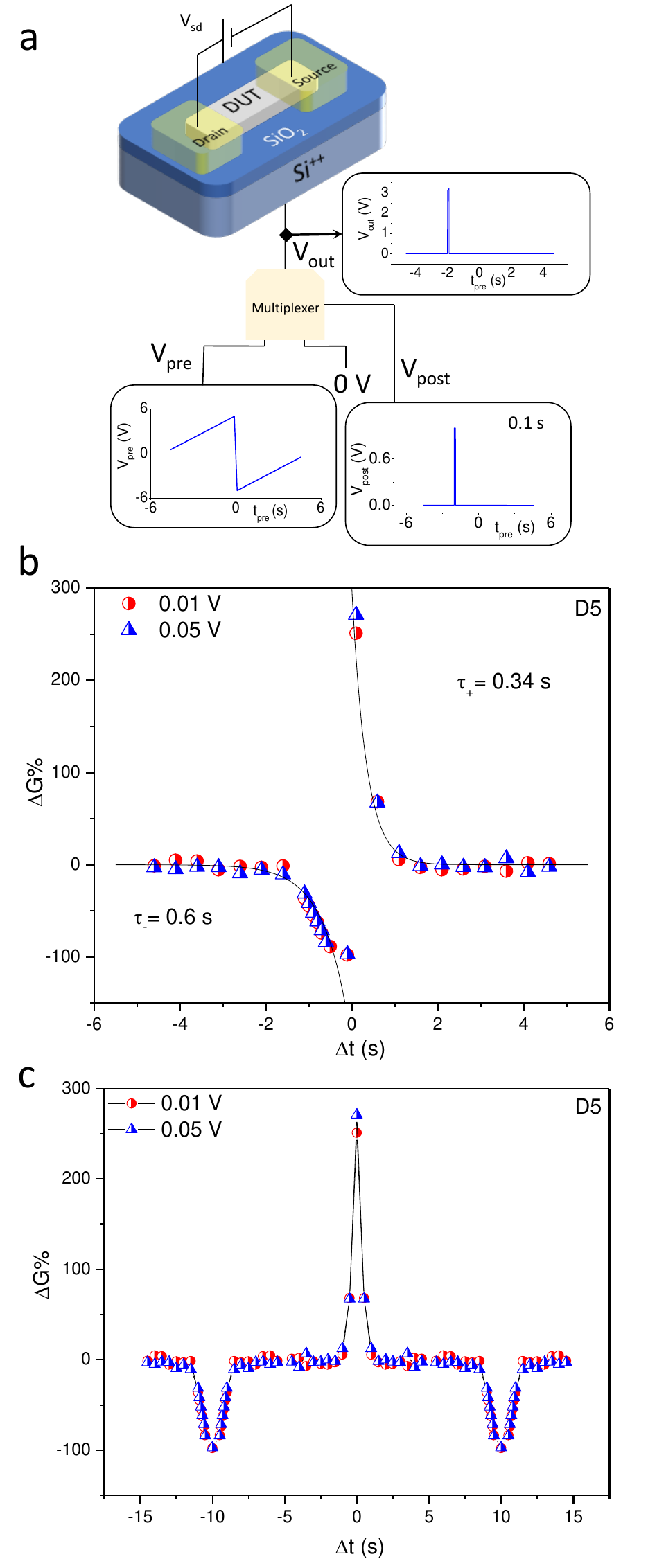}
\caption{Demonstration of synaptic plasticity. (a) Schematic depiction of the circuit used for performing spike time dependent plasticity (STDP) experiments. Demonstration of asymmetric (b) and symmetric (c) spike time dependent plasticity in MoS$_2$ based synaptic transistors. Black lines in (b) are exponential fits to asymmetric STDP plots. }
\end{figure}

Apart from the systematic modification of channel conductance in response to pre-synaptic pulsing, synaptic memories are also meant to follow specific learning mechanisms which guide their response to a train of applied pulses. Here, we demonstrate a very common learning process of the human brain known as spike time dependent plasticity (STDP) using the current device~\cite{bi_synaptic,song_competitive,froemke_spike}. In this case, the conductivity of the synapse is a function of the time difference between the pre and post synaptic pulses. This is performed using a mapping function which converts the time difference between the pulses to the magnitude of pre-synaptic pulse applied. The experimental procedure followed is demonstrated in Figure~4(a) and is similar to the process detailed in Ref.~\citen{shi_correlated}~(see Supplementary material section \RNum{5} \& \RNum{6} for more details). Depending on the mapping function used (details provided in Supplementary section \RNum{6}), we obtain synaptic responses which are symmetric (symmetric STDP)~(Figure~4(c)) or asymmetric (asymmetric STDP)~(Figure~4(b)) with respect to the time difference between the pre and post synaptic  pulses.  To demonstrate the effect, we have plotted the percentage change in channel conductance $\Delta G\%$ with the time difference $\Delta t $. $\Delta G\%$ is given by 
\begin{figure}
\includegraphics[width=1\linewidth]{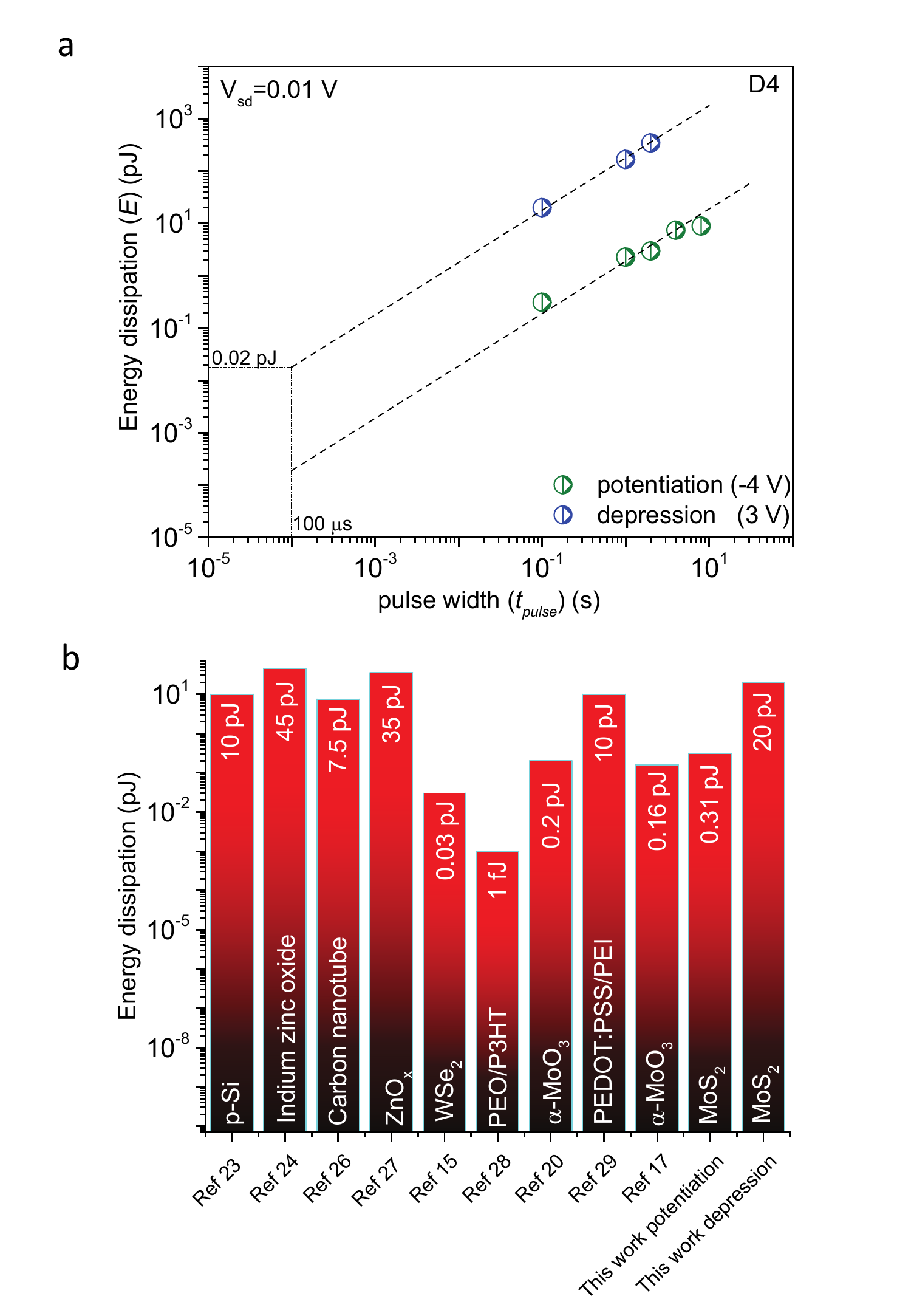}
\caption{Energy dissipation in MoS$_2$ FG devices. (a) Energy dissipation (Eq.~\ref{energy}) vs pulse width for potentiation and depression pulses. Dashed lines are linear fits to the observed dissipation. (b) Comparison of energy consumption per pulse in the current device with various other synaptic transistors reported till date. The active element and energy consumed per pulse for all the references used in the comparison can be found at the bottom and top of each individual bar respectively. }
\end{figure}

\begin{equation}
\label{percentage change in conductance}
   \Delta G\% =  \frac{G_{final}-G_{initial}}{G_{initial}}\times100\%
\end{equation}
\noindent
where $G_{intial}$ and $G_{final}$ are the channel conductance before and after the application of the synaptic pulse respectively. We observe large changes in the synaptic weight for small time differences between the pre and post synaptic pulse in both types of synaptic learning (Figure~4(b) and (c)). For the asymmetric case (Figure~4(b)), we see a sharp decrease in channel conductance for a non causal event, $i.e.$ $\Delta t\less 0$, while there is a sharp increase in conductivity for a causal event, $i.e.$ $\Delta t\ge 0$. To obtain a time constant for the potentiation and depression pulses we fit an exponential to the STDP data in Figure~4(b) (black solid lines) as follows~\cite{song_competitive}
\begin{equation}
\label{stdp}
   \Delta G \propto 
\begin{cases}
    \exp(-\frac{\Delta t}{\tau_+}),& \text{if } \Delta t\geq 0\\
    -\exp(\frac{\Delta t}{\tau_-}) , & \text{if } \Delta t\leq 0\\
\end{cases}
\end{equation}
$\tau_+$ and $\tau_-$ denote the characteristic scale of time difference between the pre and post synaptic pulses for which there is a considerable change in the synaptic weight. For the current device we find these values to be 0.34~s and 0.6~s for potentiation and depression pulses respectively. These values can be tuned by changing the mapping function (Supplementary section \RNum{6}). For the symmetric STDP case (Figure~4(c)) we find that the channel conductivity depends only on the absolute time difference  between the synaptic inputs $\mid\Delta t\mid$. The change in channel conductivity is $\approx$ 100$\%$ leading to a very robust demonstration of spike time dependent learning which is independent of the applied bias (Figure~4(b) and (c)).

To evaluate the energy efficiency of the our synaptic transistor, note that the energy dissipated for a single pulse is given by
\begin{equation}
\label{energy}
   E= I_{sd}\times t_{pulse}\times V_{sd}
\end{equation}
where $I_{sd}$ is the average current during the pulse,  $t_{pulse}$ is the time period of the pulse and $V_{sd}$ the drain bias. Figure~5(a) plots the energy dissipation as a function of pulse width for both potentiation ($-4$~V pulse height) and depression ($+3$~V pulse height) pulses at a drain bias ($V_{sd}$) of 0.01~V for the synaptic device D4. Since the channel conductance is lower during a potentiation pulse and higher during a depression pulse, we observe a higher energy loss during depression (Figure~5(a)). The observed energy dissipation $\approx$~20 pJ per pulse for depression is similar to synaptic devices previously reported~\cite{lai_ionic,zhu_artificial,gkoupidenis_neuromorphic,kim_carbon,balakrishna_nanoionics,xu_organic,van_non,yan_graphene,yang_all,yang_synaptic,zhu_ion} (Figure~5(b)). Notably, this is about five decades lower than similar devices operated in two terminal geometry ($\approx$ 1~$\mu$J per pulse for same pulse duration)~\cite{vu_two} and $\sim 1 - 2$ decades lower than complementary MOS devices~\cite{indiveri_vlsi,yang_synaptic}. We also note that the energy dissipation in our devices scale linearly with pulse width (Figure~5(a)) leading to a decrease in energy consumption for lower values of $t_{pulse}$ (Eq.~\ref{energy}). For our MoS$_2$ based synaptic transistor, we find that the extrapolated energy dissipation for a pulse width of $\approx$ 100 $\mu$s is $\approx$ 20~fJ (indicated in Figure~5(a)), which is comparable to that in Ref.~\citen{zhu_ion}, reiterating the benefits of using TMDC based synaptic transistors for enhanced power efficiency. Additionally, we now know that both in-plane and cross-plane charge and heat transport in van der Waals heterostructures are strongly temperature dependent and can be tuned accurately with external electric fields.~\cite{koppens_out,Kallol_Gr_MoS2_hybrid_Nat_Nano}  This allows a holistic integration of transport layer, heating layer and floating gate in a bottom up fashion,~\cite{mahapatra_seebeck} opening up a wide range of possibilities including the implementation of biorealistic neuromorphic realizations for example, by electro-thermal pulsing in a second order memristor.~\cite{kim_experimental}

In conclusion, we have successfully fabricated a charge-tunneling based synaptic transistor using ultrathin molybdenum disulphide channels. Repeated potentiation and depression of the channel conductance is demonstrated along with spike timing dependent synaptic plasticity while maintaining a desirable energy efficiency. We provide a new framework for solid state  synaptic devices free of electrochemical reactions which may be utilised in future neuromorphic applications.

\medskip

\section*{Acknowledgement}

We acknowledge the Department of Science and Technology (DST) for a funded project. The authors would also like to thank National Nanofabrication Facility (NNFC), CENSE, IISC and Micro and Nano Characterization Facility (MNCF), CENSE, IISC for fabrication and characterization facilities provided.


\begin{thebibliography}{10}
\providecommand{\url}[1]{\texttt{#1}}
\providecommand{\urlprefix}{URL }

\bibitem{hebb_book}
D.~Hebb.
\newblock \emph{The Organization of Behavior: A Neuropsychological Theory}.
\newblock Taylor \& Francis, \textbf{2002}.

\bibitem{guillery_binocular}
R.~Guillery.
\newblock \emph{J. Comp. Neurol.} \textbf{1972}, \emph{144}, 1 117.

\bibitem{miller_synaptic}
K.~D. Miller.
\newblock \emph{Neuron} \textbf{1996}, \emph{17}, 3 371.

\bibitem{Neumann_computation}
A.~Burks, H.~Goldstein, J.~Von~Neumann.
\newblock \emph{Logical Design of an Electronic Computing Instrument}.
\newblock Princeton, \textbf{1946}.

\bibitem{hennessy_book}
J.~L. Hennessy, D.~A. Patterson.
\newblock \emph{Computer architecture: a quantitative approach}.
\newblock Elsevier, \textbf{2011}.

\bibitem{Mead_neuromorphic}
C.~Mead.
\newblock \emph{Proc. IEEE} \textbf{1990}, \emph{78}, 10 1629.

\bibitem{thakur_large}
C.~S. Thakur, J.~Molin, G.~Cauwenberghs, G.~Indiveri, K.~Kumar, N.~Qiao,
  J.~Schemmel, R.~Wang, E.~Chicca, J.~O. Hasler, J.~Seo, S.~Yu, Y.~Cao, A.~van
  Schaik, R.~Etienne{-}Cummings.
\newblock \emph{arXiv preprint arXiv:1805.08932} \textbf{2018}.

\bibitem{thakur_analogue}
C.~S. Thakur, R.~Wang, T.~J. Hamilton, R.~Etienne-Cummings, J.~Tapson, A.~van
  Schaik.
\newblock \emph{{IEEE} Trans. Circuits Syst. {I}} \textbf{2018}, \emph{65}, 4
  1174.

\bibitem{thakur_neuromorphic}
C.~S. Thakur, T.~J. Hamilton, R.~Wang, J.~Tapson, A.~van Schaik.
\newblock In \emph{Neural Networks (IJCNN), 2015 International Joint Conference
  on}. IEEE, \textbf{2015} 1--8.

\bibitem{thakur_sound}
C.~S. Thakur, R.~M. Wang, S.~Afshar, T.~J. Hamilton, J.~Tapson, S.~Shamma,
  A.~van Schaik.
\newblock \emph{Front. Neurosci.} \textbf{2015}, \emph{9} 309.

\bibitem{young_short}
K.~K. Young.
\newblock \emph{IIEEE Trans. Electron Devices} \textbf{1989}, \emph{36}, 2 399.

\bibitem{desai_mos2}
S.~B. Desai, S.~R. Madhvapathy, A.~B. Sachid, J.~P. Llinas, Q.~Wang, G.~H. Ahn,
  G.~Pitner, M.~J. Kim, J.~Bokor, C.~Hu, H.-S.~P. Wong, A.~Javey.
\newblock \emph{Science} \textbf{2016}, \emph{354}, 6308 99.

\bibitem{Peide_Ye_Channel_length_scaling}
H.~Liu, A.~T. Neal, P.~D. Ye.
\newblock \emph{ACS Nano} \textbf{2012}, \emph{6}, 10 8563.

\bibitem{lodish_neurotransmitters}
H.~Lodish, A.~Berk, S.~L. Zipursky, P.~Matsudaira, D.~Baltimore, J.~Darnell.
\newblock \emph{Neurotransmitters, synapses, and impulse transmission}.
\newblock WH Freeman, \textbf{2000}.

\bibitem{zhu_ion}
J.~Zhu, Y.~Yang, R.~Jia, Z.~Liang, W.~Zhu, Z.~U. Rehman, L.~Bao, X.~Zhang,
  Y.~Cai, L.~Song, R.~Huang.
\newblock \emph{Adv. Mater.} \textbf{2018}, \emph{30}, 21 1800195.

\bibitem{shi_correlated}
J.~Shi, S.~D. Ha, Y.~Zhou, F.~Schoofs, S.~Ramanathan.
\newblock \emph{Nat. Commun.} \textbf{2013}, \emph{4} 2676.

\bibitem{yang_all}
C.-S. Yang, D.-S. Shang, N.~Liu, E.~J. Fuller, S.~Agrawal, A.~A. Talin, Y.-Q.
  Li, B.-G. Shen, Y.~Sun.
\newblock \emph{Adv. Funct. Mater.} \textbf{2018}, 1804170.

\bibitem{sangwan_multi}
V.~K. Sangwan, H.-S. Lee, H.~Bergeron, I.~Balla, M.~E. Beck, K.-S. Chen, M.~C.
  Hersam.
\newblock \emph{Nature} \textbf{2018}, \emph{554}, 7693 500.

\bibitem{tian_anisotropic}
H.~Tian, Q.~Guo, Y.~Xie, H.~Zhao, C.~Li, J.~J. Cha, F.~Xia, H.~Wang.
\newblock \emph{Adv. Mater.} \textbf{2016}, \emph{28}, 25 4991.

\bibitem{yang_synaptic}
C.~S. Yang, D.~S. Shang, N.~Liu, G.~Shi, X.~Shen, R.~C. Yu, Y.~Q. Li, Y.~Sun.
\newblock \emph{Adv. Mater.} \textbf{2017}, \emph{29}, 27 1700906.

\bibitem{chang_short}
T.~Chang, S.-H. Jo, W.~Lu.
\newblock \emph{ACS Nano} \textbf{2011}, \emph{5}, 9 7669.

\bibitem{jo_nanoscale}
S.~H. Jo, T.~Chang, I.~Ebong, B.~B. Bhadviya, P.~Mazumder, W.~Lu.
\newblock \emph{Nano Lett.} \textbf{2010}, \emph{10}, 4 1297.

\bibitem{lai_ionic}
Q.~Lai, L.~Zhang, Z.~Li, W.~F. Stickle, R.~S. Williams, Y.~Chen.
\newblock \emph{Adv. Mater.} \textbf{2010}, \emph{22}, 22 2448.

\bibitem{zhu_artificial}
L.~Q. Zhu, C.~J. Wan, L.~Q. Guo, Y.~Shi, Q.~Wan.
\newblock \emph{Nat. Commun.} \textbf{2014}, \emph{5} 3158.

\bibitem{gkoupidenis_neuromorphic}
P.~Gkoupidenis, N.~Schaefer, B.~Garlan, G.~G. Malliaras.
\newblock \emph{Adv. Mater.} \textbf{2015}, \emph{27}, 44 7176.

\bibitem{kim_carbon}
K.~Kim, C.-L. Chen, Q.~Truong, A.~M. Shen, Y.~Chen.
\newblock \emph{Adv. Mater.} \textbf{2013}, \emph{25}, 12 1693.

\bibitem{balakrishna_nanoionics}
P.~Balakrishna~Pillai, M.~M. De~Souza.
\newblock \emph{ACS Appl. Mater. Interfaces} \textbf{2017}, \emph{9}, 2 1609.

\bibitem{xu_organic}
W.~Xu, S.-Y. Min, H.~Hwang, T.-W. Lee.
\newblock \emph{Sci. Adv.} \textbf{2016}, \emph{2}, 6 e1501326.

\bibitem{van_non}
Y.~van~de Burgt, E.~Lubberman, E.~J. Fuller, S.~T. Keene, G.~C. Faria,
  S.~Agarwal, M.~J. Marinella, A.~A. Talin, A.~Salleo.
\newblock \emph{Nat. Mater.} \textbf{2017}, \emph{16}, 4 414.

\bibitem{yan_graphene}
X.~Yan, L.~Zhang, H.~Chen, X.~Li, J.~Wang, Q.~Liu, C.~Lu, J.~Chen, H.~Wu,
  P.~Zhou.
\newblock \emph{Adv. Funct. Mater.} \textbf{2018}, 1803728.

\bibitem{Philip_Kim_controlled_charge_trapping}
M.~Sup~Choi, G.-H. Lee, Y.-J. Yu, D.-Y. Lee, S.~Hwan~Lee, P.~Kim, J.~Hone,
  W.~Jong~Yoo.
\newblock \emph{Nat. Commun.} \textbf{2013}, \emph{4} 1624.

\bibitem{Andras_kis_MoS2_memory}
S.~Bertolazzi, D.~Krasnozhon, A.~Kis.
\newblock \emph{ACS Nano} \textbf{2013}, \emph{7}, 4 3246.

\bibitem{Seongil_MoS2_memory}
H.~S. Lee, S.-W. Min, M.~K. Park, Y.~T. Lee, P.~J. Jeon, J.~H. Kim, S.~Ryu,
  S.~Im.
\newblock \emph{Small} \textbf{2012}, \emph{8}, 20 3111.

\bibitem{Choi_Mos2_memory}
M.~H. Woo, B.~C. Jang, J.~Choi, K.~J. Lee, G.~H. Shin, H.~Seong, S.~G. Im,
  S.-Y. Choi.
\newblock \emph{Adv. Funct. Mater.} \textbf{2017}, \emph{27}, 43 1703545.

\bibitem{Andras_Kis_Nature_Nano_review}
Q.~H. Wang, K.~Kalantar-Zadeh, A.~Kis, J.~N. Coleman, M.~S. Strano.
\newblock \emph{Nat. Nanotechnol.} \textbf{2012}, \emph{7} 699.

\bibitem{Jar_TMDC_review}
D.~Jariwala, V.~K. Sangwan, L.~J. Lauhon, T.~J. Marks, M.~C. Hersam.
\newblock \emph{ACS Nano} \textbf{2014}, \emph{8}, 2 1102.

\bibitem{Kallol_Gr_MoS2_hybrid_Nat_Nano}
K.~Roy, M.~Padmanabhan, S.~Goswami, T.~Sai, G.~Ramalingam, S.~Raghavan,
  A.~Ghosh.
\newblock \emph{Nat. Nanotechnol.} \textbf{2013}, \emph{8} 826.

\bibitem{Subhamoy_ACS_Nano}
S.~Ghatak, A.~N. Pal, A.~Ghosh.
\newblock \emph{ACS Nano} \textbf{2011}, \emph{5}, 10 7707.

\bibitem{Percolative_switching}
T.~Paul, S.~Ghatak, A.~Ghosh.
\newblock \emph{Nanotechnol.} \textbf{2016}, \emph{27}, 12 125706.

\bibitem{Andras_Kis_MoS2_transistor}
B.~Radisavljevic, A.~Radenovic, J.~Brivio, V.~Giacometti, A.~Kis.
\newblock \emph{Nat. Nanotechnol.} \textbf{2011}, \emph{6}, 3 147.

\bibitem{Dong_Sun_Valley}
Q.~Wang, S.~Ge, X.~Li, J.~Qiu, Y.~Ji, J.~Feng, D.~Sun.
\newblock \emph{ACS Nano} \textbf{2013}, \emph{7}, 12 11087.

\bibitem{Ji_Feng_Valley}
T.~Cao, G.~Wang, W.~Han, H.~Ye, C.~Zhu, J.~Shi, Q.~Niu, P.~Tan, E.~Wang,
  B.~Liu, J.~Feng.
\newblock \emph{Nat. Commun.} \textbf{2012}, \emph{3}, 9 887.

\bibitem{Tony_F_Heinz_Valley}
K.~F. Mak, K.~He, J.~Shan, T.~F. Heinz.
\newblock \emph{Nat. Nanotechnol.} \textbf{2012}, \emph{7}, 8 494.

\bibitem{cui_valley}
H.~Zeng, J.~Dai, W.~Yao, D.~Xiao, X.~Cui.
\newblock \emph{Nat. Nanotechnol.} \textbf{2012}, \emph{7} 490.

\bibitem{Coupled_spin_valley_Wang}
D.~Xiao, G.-B. Liu, W.~Feng, X.~Xu, W.~Yao.
\newblock \emph{Phys. Rev. Lett.} \textbf{2012}, \emph{108} 196802.

\bibitem{vu_two}
Q.~A. Vu, Y.~S. Shin, Y.~R. Kim, W.~T. Kang, H.~Kim, D.~H. Luong, I.~M. Lee,
  K.~Lee, D.-S. Ko, J.~Heo, Y.~H. Lee, W.~J. Yu.
\newblock \emph{Nat. Commun.} \textbf{2016}, \emph{7} 12725.

\bibitem{wang_floating}
J.~Wang, X.~Zou, X.~Xiao, L.~Xu, C.~Wang, C.~Jiang, J.~C. Ho, T.~Wang, J.~Li,
  L.~Liao.
\newblock \emph{Small} \textbf{2015}, \emph{11}, 2 208.

\bibitem{frohman_electrically}
D.~Frohman-Bentchkowsky, J.~Mar, G.~Perlegos, W.~S. Johnson.
\newblock Electrically programmable and erasable mos floating gate memory
  device employing tunneling and method of fabricating same, \textbf{1980}.
\newblock US Patent 4,203,158.

\bibitem{diorio_single}
C.~Diorio, P.~Hasler, A.~Minch, C.~A. Mead.
\newblock \emph{IEEE Trans. Electron Devices} \textbf{1996}, \emph{43}, 11
  1972.

\bibitem{geim_van}
A.~K. Geim, I.~V. Grigorieva.
\newblock \emph{Nature} \textbf{2013}, \emph{499}, 7459 419.

\bibitem{aamir20172d}
M.~A. Aamir, T.~Ahmed, K.~Hsieh, S.~Islam, P.~Karnatak, R.~Kashid, P.~S.
  Mahapatra, J.~Mishra, T.~Paul, A.~Pradhan, K.~Roy, A.~Sahoo, A.~Ghosh.
\newblock \emph{2D van der Waals Hybrid: Structures, Properties and Devices}.
\newblock World Scientific, \textbf{2017}.

\bibitem{Watanabe_hbn_bandgap}
K.~Watanabe, T.~Taniguchi.
\newblock \emph{Nat. Mater.} \textbf{2004}, \emph{3} 404.

\bibitem{Subhamoy_APL_Mat}
S.~Ghatak, S.~Mukherjee, M.~Jain, D.~D. Sarma, A.~Ghosh.
\newblock \emph{APL Mat.} \textbf{2014}, \emph{2}, 9 092515.

\bibitem{Dean_graphene_mobility}
C.~R. Dean, A.~F. Young, I.~Meric, C.~Lee, L.~Wang, S.~Sorgenfrei, K.~Watanabe,
  T.~Taniguchi, P.~Kim, K.~L. Shepard, J.~Hone.
\newblock \emph{Nat. Nanotechnol.} \textbf{2010}, \emph{5} 722.

\bibitem{pari_noise_review}
P.~Karnatak, T.~Paul, S.~Islam, A.~Ghosh.
\newblock \emph{Advances in Physics: X} \textbf{2017}, \emph{2}, 2 428.

\bibitem{Pari_current_crowding}
P.~Karnatak, T.~P. Sai, S.~Goswami, S.~Ghatak, S.~Kaushal, A.~Ghosh.
\newblock \emph{Nat. Commun.} \textbf{2016}, \emph{7} 13703.

\bibitem{Sakurai_alpha_power_law}
T.~Sakurai, A.~R. Newton.
\newblock \emph{IEEE J. Solid-State Circuits} \textbf{1990}, \emph{25}, 2 584.

\bibitem{Howritz_low_power_CMOS}
R.~Gonzalez, B.~M. Gordon, M.~A. Horowitz.
\newblock \emph{IEEE J. Solid-State Circuits} \textbf{1997}, \emph{32}, 8 1210.

\bibitem{Martin_transrerional_model}
S.~Keller, D.~M. Harris, A.~J. Martin.
\newblock \emph{IEEE Trans. Very Large Scale Integr. (VLSI) Syst.}
  \textbf{2014}, \emph{22}, 10 2041.

\bibitem{lee_electron}
G.-H. Lee, Y.-J. Yu, C.~Lee, C.~Dean, K.~L. Shepard, P.~Kim, J.~Hone.
\newblock \emph{Appl. Phys. Lett.} \textbf{2011}, \emph{99}, 24 243114.

\bibitem{bi_synaptic}
G.-q. Bi, M.-m. Poo.
\newblock \emph{J. Neurosci.} \textbf{1998}, \emph{18}, 24 10464.

\bibitem{song_competitive}
S.~Song, K.~D. Miller, L.~F. Abbott.
\newblock \emph{Nat. Neurosci.} \textbf{2000}, \emph{3}, 9 919.

\bibitem{froemke_spike}
R.~C. Froemke, Y.~Dan.
\newblock \emph{Nature} \textbf{2002}, \emph{416}, 6879 433.

\bibitem{indiveri_vlsi}
G.~Indiveri, E.~Chicca, R.~J. Douglas.
\newblock \emph{IEEE Trans. Neural Networks} \textbf{2006}, \emph{17}, 1.

\bibitem{koppens_out}
K.-J. Tielrooij, N.~C. Hesp, A.~Principi, M.~B. Lundeberg, E.~A. Pogna,
  L.~Banszerus, Z.~Mics, M.~Massicotte, P.~Schmidt, D.~Davydovskaya, et~al.
\newblock \emph{Nat. Nanotechnol.} \textbf{2018}, \emph{13}, 1 41.

\bibitem{mahapatra_seebeck}
P.~S. Mahapatra, K.~Sarkar, H.~R. Krishnamurthy, S.~Mukerjee, A.~Ghosh.
\newblock \emph{Nano Lett.} \textbf{2017}, \emph{17}, 11 6822.

\bibitem{kim_experimental}
S.~Kim, C.~Du, P.~Sheridan, W.~Ma, S.~Choi, W.~D. Lu.
\newblock \emph{Nano Lett.} \textbf{2015}, \emph{15}, 3 2203.

\end{thebibliography}

\end{document}


\pagenumbering{gobble}
\beginsupplement
\title{Supplementary: A high-performance MoS$_2$ synaptic device with floating gate engineering for Neuromorphic Computing}

\author{Tathagata Paul,$^1$\footnotemark[3] Tanweer Ahmed,$^1$ Krishna Kanhaiya Tiwari,$^2$ Chetan Singh Thakur$^3$ and Arindam Ghosh$^{1,4}$\footnote[3]{e-mail:tathagata@iisc.ac.in, arindam@iisc.ac.in}}

\address{$^1$Department of Physics, Indian Institute of Science, Bangalore 560012,India.  $^2$Visva Bharati University Santiniketan, West Bengal 731235, India. $^3$Department of Electronic Systems Engineering, Indian Institute of Science, Bangalore 560012, India. $^4$Centre for Nanoscience and Engineering Indian Institute of Science,Bangalore 560012, India.}

\maketitle
\newpage
\section{D\MakeLowercase{evices} \MakeLowercase{used}}

\begin{table}[h]
\renewcommand{\arraystretch}{2}
\renewcommand{\tabcolsep}{5mm}
\begin{center}
\caption{Details of various MoS$_2$ FG devices fabricated}
\label{Devices}
\begin{tabular}{c|c|c|c }
 \Xhline{2\arrayrulewidth}
  \Xhline{2\arrayrulewidth}
\textbf{Device name} & \textbf{thickness of hBN (nm)}  & \textbf{channel length ($\mu$m)} & \textbf{channel width ($\mu$m)} \\
    \Xhline{2\arrayrulewidth}
    \Xhline{2\arrayrulewidth}

    D1 & 5.9 & 0.85 & 0.82 \\
    \hline

    D2 & 6.4 & 0.65 & 1.05 \\
    \hline

    D3 & 6.7 & 0.43 & 0.47 \\
    \hline

    D4 & 5.8 & 0.53 & 0.95 \\
    \hline
    
    D5 & 5.8 & 0.6 & 0.95 \\
    \hline
    
    D6 & 6 & 0.9 & 0.47 \\
    \hline
    
    D7 & 7 & 0.75 & 0.83 \\
    \hline
    
    D8 & 8.5 & 0.8 & 1 \\
    \hline
    
    D9 (without extended FG) & 7.5 & 0.8 & 2.3 \\
    \hline
    
    D10 (without FG) & 4 & 1 & 1.1 \\
    \hline
\end{tabular}
\end{center}
\renewcommand{\arraystretch}{1}
\end{table}

\section{S\MakeLowercase{ubthreshold} S\MakeLowercase{wing}} 
One of the key factors limiting the performance of FETs is the subthreshold swing. It is given by inverse of the amount of gate bias required to change the drain current by one decade and generally determines how fast a transistor switches. The subthreshold swing ($S$) maybe represented as
\begin{equation}
\label{subthreshold_swing}
S = \frac{\partial{V_g}}{\partial{({\rm log_{10}}I_{sd})}}=\frac{\partial{V_g}}{\partial{\psi_s}}\frac{\partial{\psi_s}}{\partial{({\rm log_{10}}I_{sd})}}
\end{equation}
where ${V_g}$ is the gate bias, $I_{sd}$ the drain current and $\psi_s$ the surface potential in the channel. The second term $\frac{\partial{\psi_s}}{\partial{({\rm log_{10}}I_{sd})}}$ is theoretically pegged at a value of 60 mV/decade at room temperature while the first term $\frac{\partial{V_g}}{\partial{\psi_s}}$ also known as the body factor is given by
\begin{equation}
\label{body_factor}
\frac{\partial{V_g}}{\partial{\psi_s}}=1+\frac{C_s}{C_{eqv}}
\end{equation}

This relation arises because we can picture the capacitive gate circuit as a series combination of capacitor $C_{eqv}$ the equivalent gate capacitance and $C_s$ the surface capacitance which is the quantum capacitance of the channel. For the case of a MoS$_2$ device fabricated on SiO$_2$, $C_{eqv}$ would be the capacitance of the SiO$_2$ dielcetric. However, as observed in the main text, the equivalent capacitance for our extended floating gate structure is the  capacitance of the hBN dielectric. Due to the two dimensional nature of a hBN dielectric ( thickness $\approx$ 10~nm), the capacitance is much larger when compared to an SiO$_2$ capacitor of the same area since the thickness of the SiO$_2$ dielectric is $\approx$ 285~nm. Hence, the application of an extended floating gate effectively allows us to use the hBN dielectric as the back gate rather than the (285~nm) SiO$_2$. From Eqn.~\ref{body_factor} it is clear that higher the equivalent capacitance, lower is the value of subthrehold swing (S) leading to faster switching.  The higher capacitance per unit area of hBN allows us to reduce the body factor significantly leading to almost ideal subthreshold slope in all the measured devices for over four decades of drain current ($I_{sd}$) as shown in Fig~1f. 

\section{R\MakeLowercase{aman} \MakeLowercase{characterisation} \MakeLowercase{of} M\MakeLowercase{o}S$_2$ \MakeLowercase{and graphene}}

\begin{center}
\includegraphics[width=1\linewidth]{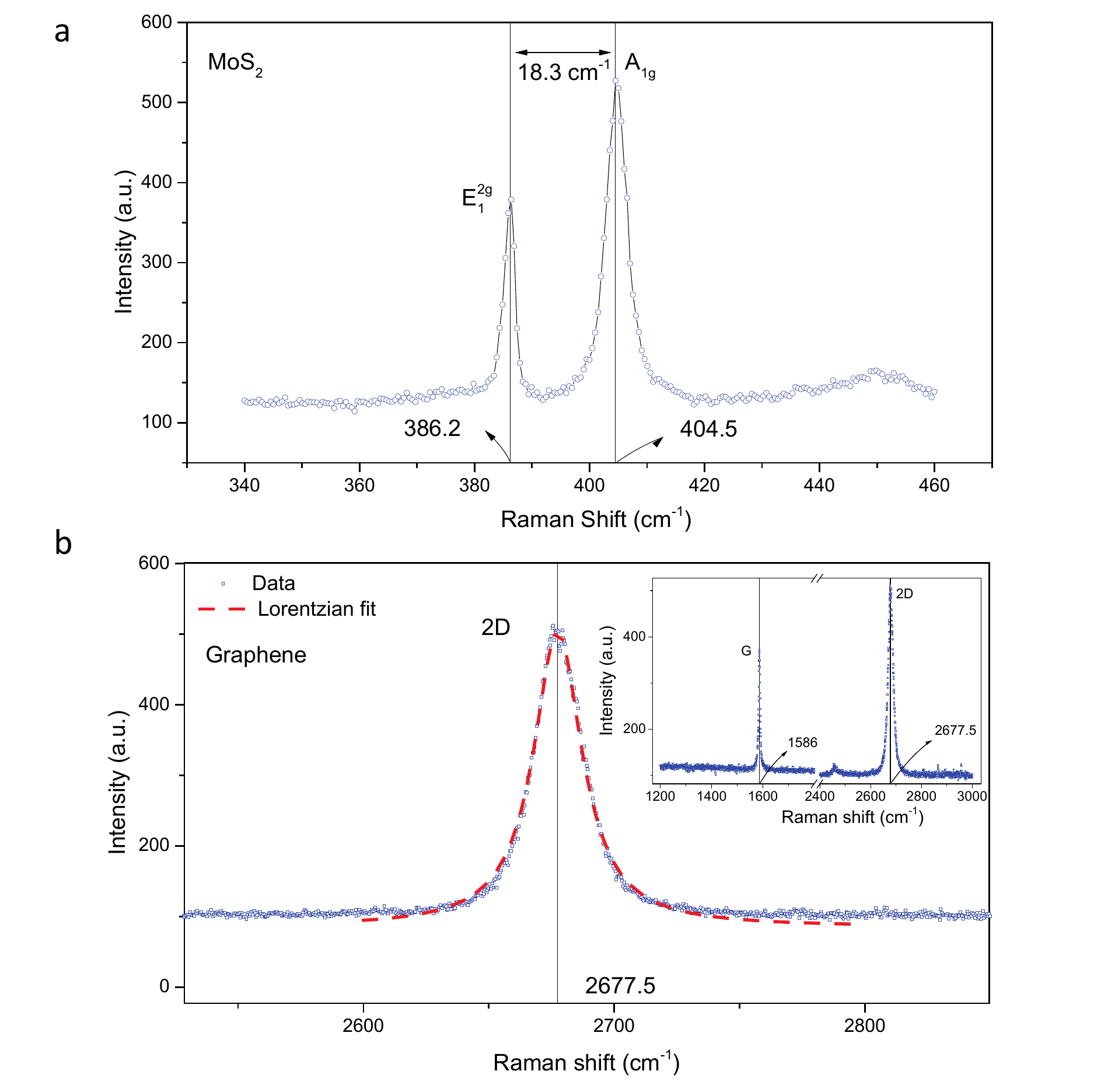}
\captionof{figure}{Raman characterisation of MoS$_2$ and graphene. (a) Raman spectra of MoS$_2$ with characteristic in-plane $E_1^{2g}$ (386 cm$^{-1}$) and out of plane $A_{1g}$ (404 cm$^{-1}$) peaks. A difference in Raman shift of $\approx$ 18 cm$^{-1}$ between these two peaks indicates a single layer MoS$_2$ flake. (b) Inset shows the Raman spectra of a graphene flake with a prominent G (1586 cm$^{-1}$) and 2D (2677 cm$^{-1}$) peaks. A single Lorentzian fit (dashed line) to the 2D peak indicates single layer graphene. Raman characterisation was performed using a 532~nm excitation laser.}
\end{center}

\section{AFM \MakeLowercase{characterisation} \MakeLowercase{of h}BN}

\begin{center}
\includegraphics[width=1\linewidth]{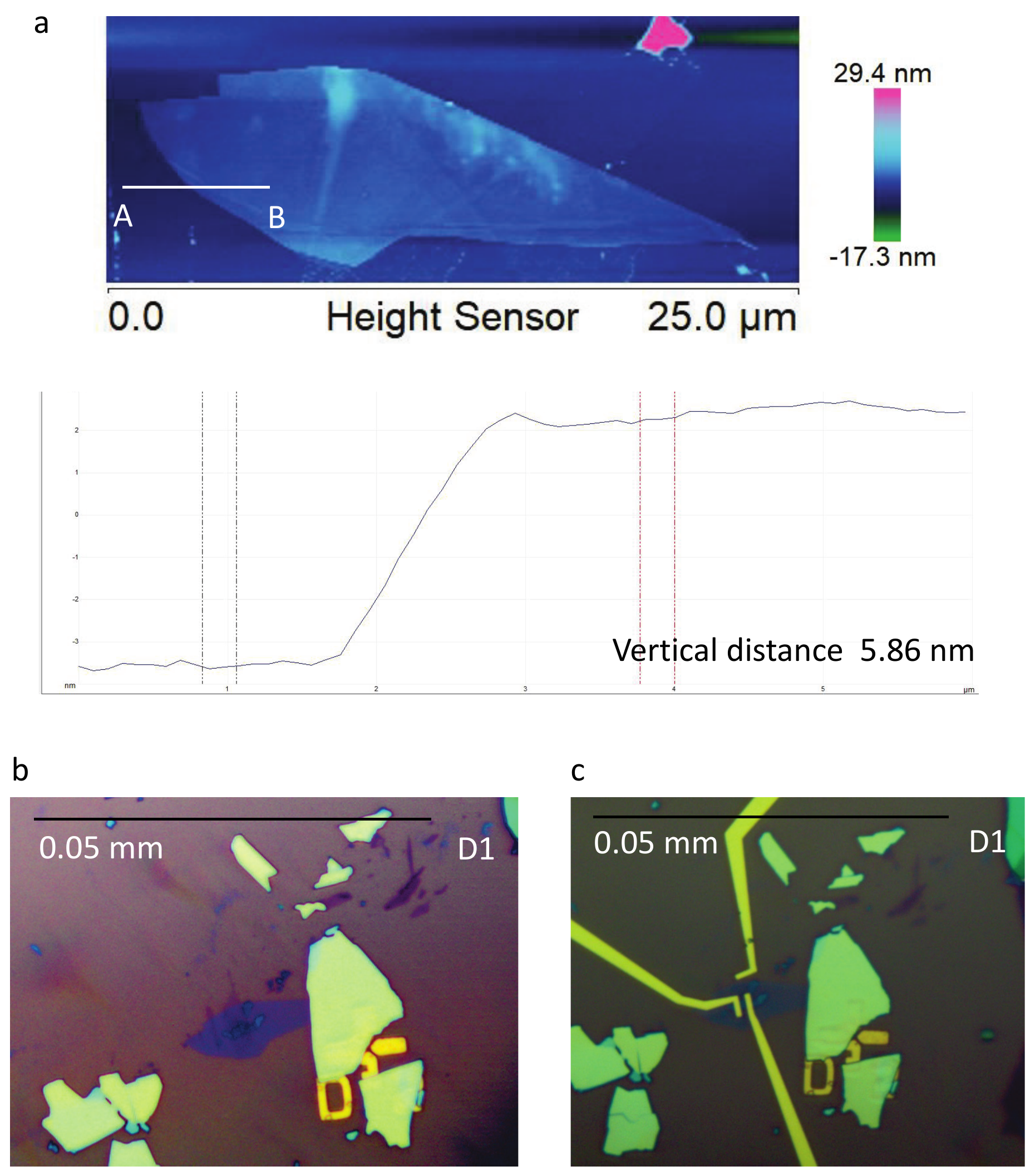}
\captionof{figure}{AFM characterisation of hBN flakes. (a) Atomic force micrograph of a typical hBN flake (top panel) with the corresponding height measurement in the bottom panel. The height measurement has been preformed along the white line `AB' in the top panel. (b) An optical micrograph of the same hBN flake in subsection (a) in a trilayer stack after transfer and (c) after electron beam lithography and metallization.  }
\end{center}

\newpage
\section{S\MakeLowercase{pike} T\MakeLowercase{ime} D\MakeLowercase{ependent} P\MakeLowercase{lasticity} (STDP) M\MakeLowercase{easurement}}

\begin{center}
\includegraphics[width=1\linewidth]{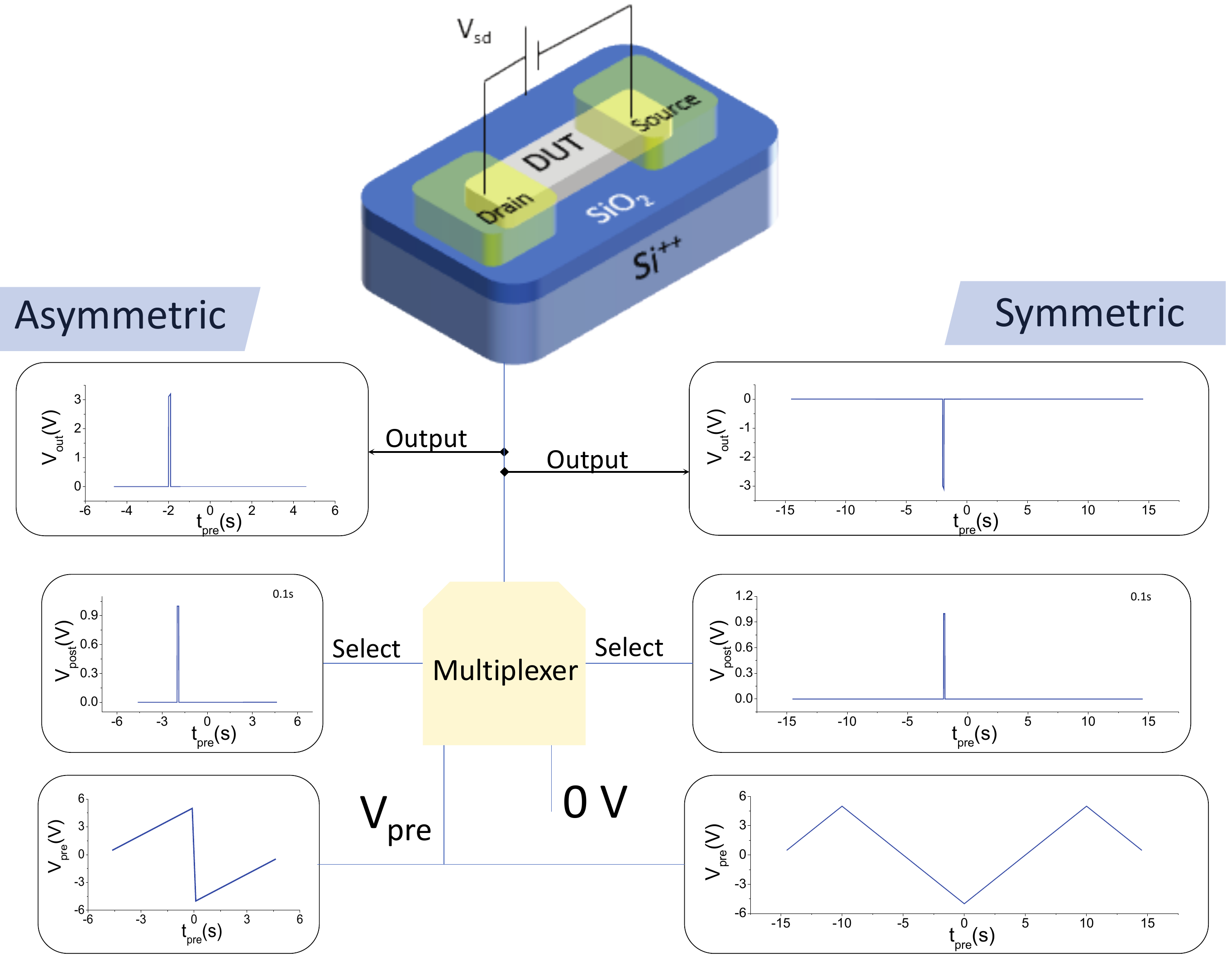}
\captionof{figure}{STDP measurement setup. Schematic of the measurement setup used for performing STDP measurements. We map the time difference between the pre and post-synaptic pulse into a voltage value using mapping functions demonstrated in the next section. This voltage value ($V_{pre}$) is fed into one of the inputs of a multiplexer while the other is held at 0~V. The select line of the multiplexer is controlled by the post synaptic pulse with the output connected to the Si$^{++}$ gate. Whenever there is a pulse in the postsynaptic channel $V_{pre}$ is applied at the gate terminal which is otherwise held at zero. Hence, the timing of the post-synaptic pulse determines the value of the gate pulse ($V_{pre}$) and consequently the change in channel conductance leading to a spike time dependent conductance change or plasticity.}
\end{center}

\section{M\MakeLowercase{apping} \MakeLowercase{functions used in} STDP}
We perform the STDP measurements by simulating time difference between the pre and post-synaptic pulse as a voltage value which is denoted by $V_{pre}$ in Fig.~S3 and is different for symmetric and asymmetric STDP. The timing of the post synaptic pulse (select pin of multiplexer) determines the pulse height of the presynaptic pulse  ($V_{pre}$) and consequently the percentage change in conductance. The mapping function used in our measurements is listed below
\newline
\newline
\textbf{Symmetric STDP}
\begin{equation}
\label{Symmetric STDP}
\begin{aligned}
V_{pre}&= \alpha \times t_{pre} + \beta \\
  \alpha& = 1~\text{V/s}\\
  \beta& = 5.1~\text{V}~\text{for}~t_{pre}<0~\text{s}\\
  \beta& = -6.1~\text{V}~\text{for}~t_{pre}>0~\text{s}
\end{aligned}
  \end{equation}
\noindent\textbf{Asymmetric STDP}
\begin{equation}
\label{Asymmetric STDP}
\begin{aligned}
V_{pre}&= \alpha \times t_{pre} + \beta \\
  \alpha& = 1~\text{V/s}~\text{for}~-15~\text{s}<t_{pre}\le -10~\text{s}~\text{and}~0~\text{s}<t_{pre}\le 10~\text{s}\\
  \alpha& = -1~\text{V/s}~\text{for}~-10~\text{s}<t_{pre}\le 0~\text{s}~\text{and}~10~\text{s}<t_{pre}\le 15~\text{s}\\
  \beta& = 15~\text{V}~\text{for}~-15~\text{s}<t_{pre}\le -10~\text{s}~\text{and}~10~\text{s}<t_{pre}\le 15~\text{s}\\
  \beta& = -5~\text{V}~\text{for}~-10~\text{s}<t_{pre}\le 0~\text{s}~\text{and}~0~\text{s}<t_{pre}\le 10~\text{s}
\end{aligned}
  \end{equation}

\section{S\MakeLowercase{weep} \MakeLowercase{rate independence of hysteresis}}

\begin{center}
\includegraphics[width=1\linewidth]{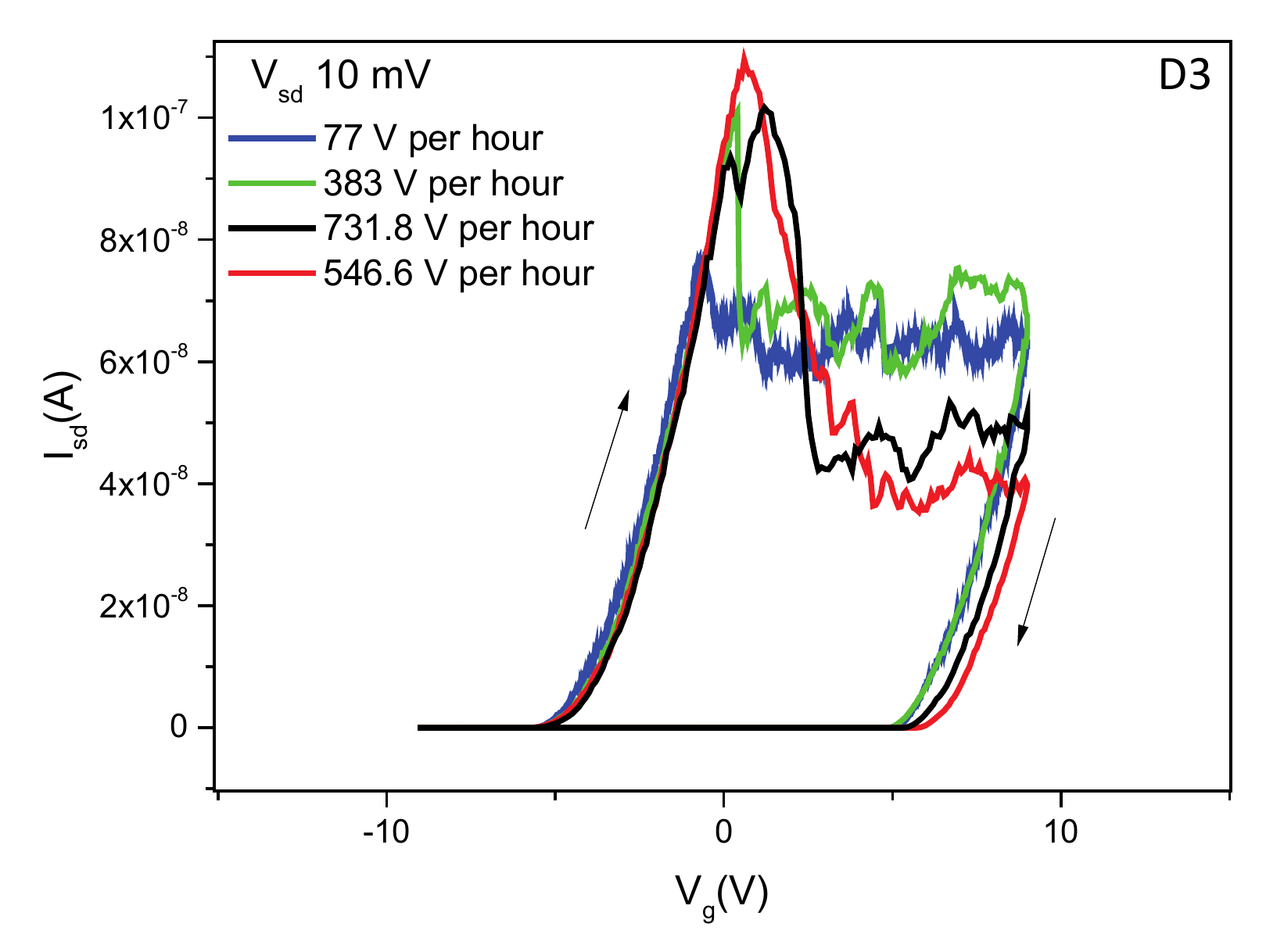}
\captionof{figure}{Sweep rate dependence of hysteresis. Transfer characteristics of a typical device performed at different sweeping rates of back gate voltage ($V_g$). A negligible change in the hysteresis window size with sweep rate indicates the absence of slow defect based charge trapping processes in the MoS$_2$ floating gate devices. The fluctuations in the ON state current is likely due to random trapping de-trapping events in the hBN or MoS2 flake.}
\end{center}

\section{P\MakeLowercase{otentiation} \MakeLowercase{and} D\MakeLowercase{epression in a two probe geometry}}
\begin{center}
\includegraphics[width=1\linewidth]{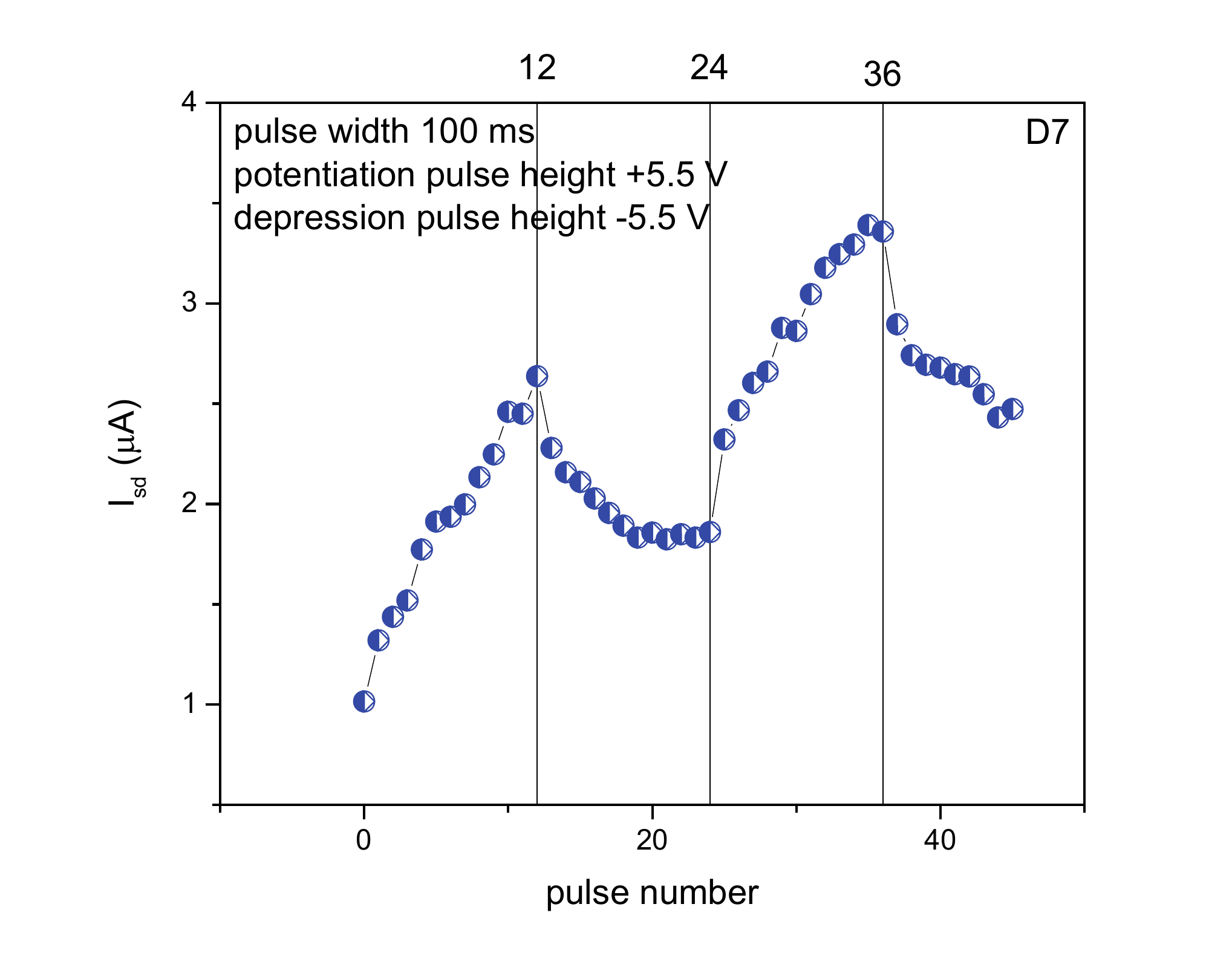}
\captionof{figure}{Synaptic Plasticity in two terminal geometry. Change in channel conductance for multiple potentiation (+ 5.5~V)and depression (-5.5~V) pulses applied to the device in a two terminal geometry. Here the pulses are applied to the drain contact and the corresponding change in conductance is measured. The readout voltage is set at $V_{sd}$ = 2~V. Though we observe potentiation and depresssion as in the three terminal geometry, the energy dissipation per pulse in this case is ($\approx$ 47~mJ) which is excessive for neuromorphic application.}
\end{center}

\section{S\MakeLowercase{tability} \MakeLowercase{of different conductance states}}
\begin{center}
\includegraphics[width=1\linewidth]{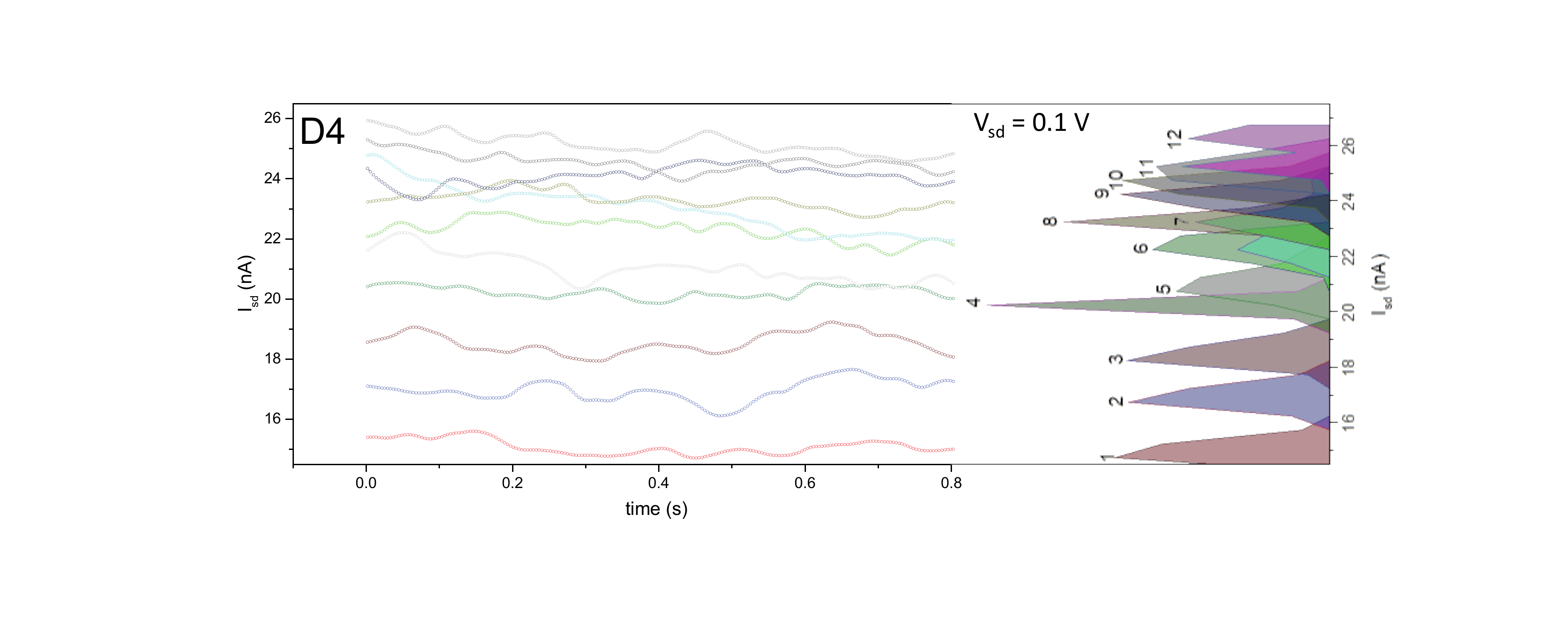}
\captionof{figure}{Stability of different conductance states. Time series for drain current in the different conductance states of the synaptic memory. The different states are accessible by applying a potentiation (-4~V) pulse. The right side shows the spread of each state in the drain current spectrum as an area plot. Different colors represent the different conductance states and are color matched with their respective time series. The peak heights are representative of the contribution of a particular value of drain current to that conductance state. The conductance states are the same ones depicted in Fig.~2b of the main text for drain bias of 0.1~V.}
\end{center}
\newpage
\section{P\MakeLowercase{aired - pulse facilitation} (PPF)}
\begin{center}
\includegraphics[width=1\linewidth]{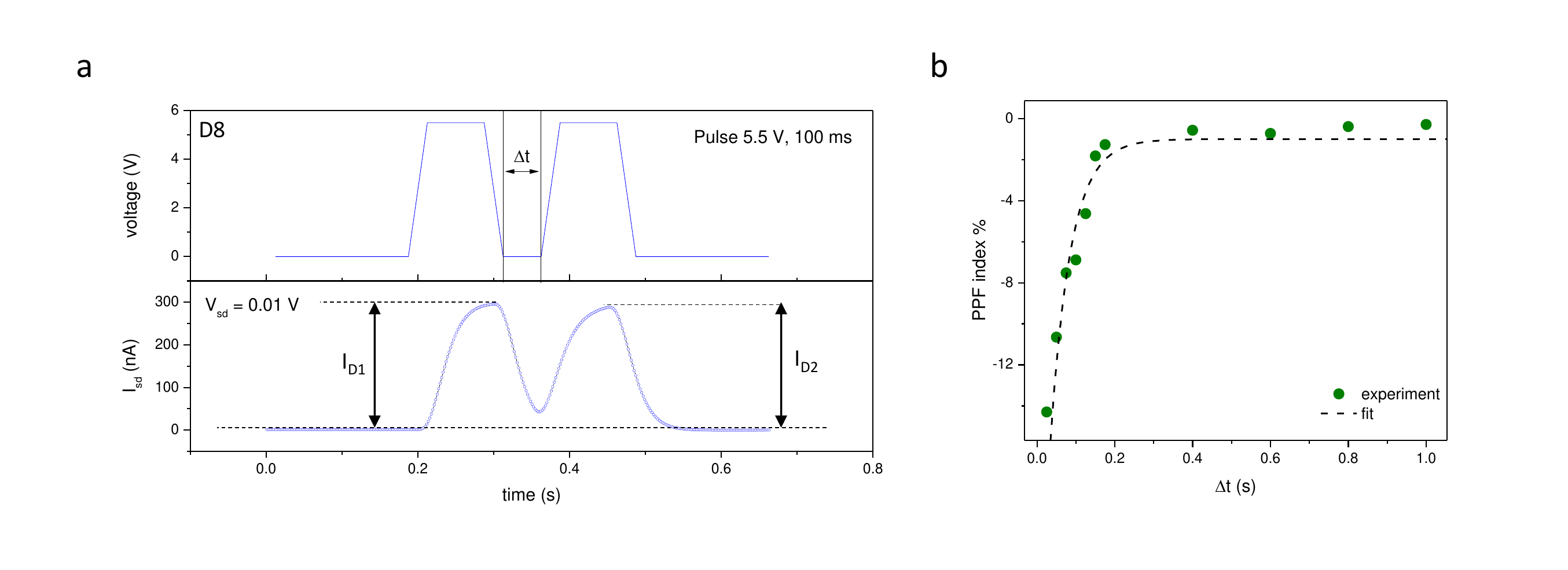}
\captionof{figure}{Short-term synaptic plasticity. (a) Top panel depicts the pulses used for demonstrating PPF while the bottom panel shows the sample response. The measurements are performed by varying the separation time ($\Delta~t$) between the pulses and observing its effect on the post synaptic current due to the second pulse. In the current device geometry, positive pulses are inhibitory pulses and hence, the post-synaptic current due to the second pulse is always lower than the first one with a stronger inhibitory action being demonstrated for smaller values of $\Delta~t$. This is quantified by computing the PPF index which denotes the percentage change in the postsynaptic current due to the second pulse and is given by, PPF index $\%$ = ($I_{D2}-I_{D1}$)/$I_{D1}$ $\%$. Subsection b demonstrates the PPF index $\%$ as a function of $\Delta~t$. The PPF index is negative, indicating an inhibitory behaviour with an exponential decrease in the inhibition strength with increasing $\Delta~t$. The black dashed line is a double exponential fit to the experimental data which is given by}
\begin{equation}
\label{PPF}
   PPF= -1- C_1exp(\frac{-t}{\tau_1})-C_2exp(\frac{-t}{\tau_2})
\end{equation}
where $t$ is the time separation between the pulses, $\tau_1$ and $\tau_2$ are the relaxation times of the two decay phases while $C_1$ and $C_2$ are the facilitation amplitudes of the respective phases. For this fitting we find, $C_1$ = 9$\%$, $C_2$ = 8$\%$, $\tau_1$ = 50~ms and $\tau_2$ = 1000~ms. The characteristic relaxation time for both phases compare well with previous reports of two dimensional system based artificial synapses.~\cite{yang_synaptic}

\end{center}